\documentclass[12pt,a4paper]{article}

\usepackage{epsfig}
\usepackage{amsmath,amsfonts,amssymb}
\usepackage{cite}
\usepackage{colortbl}
\usepackage{pifont}

\providecommand{\openone}{\leavevmode\hbox{\small1\kern-3.8pt\normalsize1}}

\newcommand{\met}{\text{MET}}

\newcommand{\esx}{\langle S_1 \rangle}
\newcommand{\esy}{\langle S_2 \rangle}

\newcommand{\esi}{\langle S_k \rangle}
\newcommand{\esz}{\langle S_3 \rangle}
\newcommand{\etz}{\langle T_0 \rangle}
\newcommand{\eax}{\langle A_1 \rangle}
\newcommand{\eay}{\langle A_2 \rangle}

\newcommand{\ebx}{\langle B_1 \rangle}
\newcommand{\eby}{\langle B_2 \rangle}

\newcommand{\eszA}{\langle S_3^A \rangle}
\newcommand{\etzA}{\langle T_0^A \rangle}

\newcommand{\ost}{\textstyle \frac{1}{\sqrt 2}}

\renewcommand{\Re}{\operatorname{Re}}
\renewcommand{\Im}{\operatorname{Im}}

\begin{document}

\begin{center}
	\begin{Large}
		{\bf The $Z$ boson spin observables \\[1mm] as messengers of new physics}
	\end{Large}

	\vspace{0.5cm}
	J.~A.~Aguilar--Saavedra$^{a}$, J.~Bernab\'eu$^{b,c}$, V.~A.~Mitsou$^c$, A. Segarra$^{b,c}$ \\[1mm]
	\begin{small}
		{$^a$ Departamento de F\'{\i}sica Te\'orica y del Cosmos, 
		Universidad de Granada, \\ E-18071 Granada, Spain} \\ 
		{$^b$ Departament de F\'{\i}sica Te\`orica, Universitat de Val\`encia, 
		E-46100 Burjassot, Spain } \\
		{$^c$ Instituto de F\'{\i}sica Corpuscular, CSIC--Universitat de Val\`encia, E-46890 Paterna, Spain } \\
	\end{small}
\end{center}

\begin{abstract}
We demonstrate that the 8 multipole parameters describing the spin state of the $Z$ boson are able to disentangle known $Z$ production mechanisms and signals from new physics at the LHC. They can be extracted from appropriate asymmetries in the angular distribution of lepton pairs from the $Z$ boson decay. The power of this analysis is illustrated by (i) the production of $Z$ boson plus jets; (ii) $Z$ boson plus missing transverse energy; (iii) $W$ and $Z$ bosons originating from the two-body decay of a heavy resonance.
\end{abstract}

\section{Introduction}

The successful operation of the Large Hadron Collider (LHC) has allowed to accumulate a wealth of collision data in the search of new physics in the ATLAS and CMS experiments, at centre-of-mass (CM) energies of 7, 8 and 13 TeV.  With the ever-increasing statistics, measurements beyond simple event counts are possible, which provide further insight into the Standard Model (SM) processes and possible new physics. Of particular interest are polarisation measurements, possible for particles with a short lifetime, through analyses of the angular distributions of their decay products. 

For spin-$1/2$ fermions there are three independent spin observables, which can be conveniently taken as 
the expectation values of the spin operators in three orthogonal directions. At the LEP experiments, this program was exploited for $\tau$ leptons~\cite{ref:LEP} and, for general $e^+ e^-$ colliders, proposed for heavy quarks~\cite{Groote:1995ky,Groote:1996nc,Tung:1996dq,Fischer:2001gp}.
But for spin-$1$ vector bosons the number of independent spin observables is eight, requiring a more elaborate discussion. There have been various studies of spin observables or decay angular distributions for vector bosons, often focusing on specific processes~\cite{Gounaris:1992kp,Mirkes:1994eb}. In previous work~\cite{Aguilar-Saavedra:2015yza} some of us have provided a full model-independent analysis of the $W$ boson spin observables. For the $Z$ boson the framework is quite similar, the main difference being that while for the $W$ boson the couplings to leptons violate parity maximally, for the $Z$ boson they do not. This difference can be taken into account by the introduction of an additional coupling factor relating the observed angular distributions to the $Z$ boson spin observables, a factor which we may call the polarisation analyser. This study has been applied to an $e^+e^-$ collider~\cite{Rahaman:2016pqj}.

The purpose of this paper is to demonstrate that the information content in the eight multipole parameters of the $Z$ boson, polarisations and alignments, is able to clearly discriminate among different production mechanisms, acting as messengers of the physics involved in the process. This is particularly important at the LHC, taking into account the hadronic environment. In Section 2 we write down the relation between $Z$ boson spin observables and the parameters of the decay angular distributions. Beyond the application to Drell-Yan $Z$ plus jets events at LHC, considered in Section 3, we move to the study of the production of a $Z$ boson plus missing transverse energy ($\met$) in Section 4, where the discrimination between the SM production mechanism and that for extended models is apparent. In Section~\ref{sec:5} we discuss in detail the predictions for the eight spin observables of $Z$ (or $W$) bosons produced from the two-body decay of a heavy resonance of spin $0$, $1/2$ or $1$. Our conclusions are presented in Section 6.

\section{Formalism}
\label{sec:2}

The spin state of $Z$ bosons is described, as for any other spin-1 massive particle, by a $3 \times 3$ density matrix $\rho$, Hermitian with unit trace and positive semidefinite. We follow closely the analysis and notation introduced~\cite{Aguilar-Saavedra:2015yza} for the analysis of the $W$ boson spin\footnote{An equivalent description of the formalism for $Z$ boson spin observables was later made in Ref.~\cite{Rahaman:2016pqj}.}. If we fix a coordinate system $(x',y',z')$ in the $Z$ boson rest frame, we can write the spin density matrix as
\begin{equation}
	\rho = \frac{1}{3} \openone + \frac{1}{2} \displaystyle \sum_{M=-1}^1 \langle S_M \rangle^* S_M + \displaystyle \sum_{M=-2}^2  \langle T_{M} \rangle^* T_{M} \,,
	\label{ec:rhoW}
\end{equation}
with $S_{\pm 1} = \mp \ost (S_1 \pm i S_2)$, $S_0 = S_3$ the spin operators in the spherical basis and $T_M$ five rank 2 irreducible tensors,
\begin{align}
	& T_{\pm 2} = S_{\pm1}^2 \,, \notag \\
	& T_{\pm 1} = \frac{1}{\sqrt 2} \left[ S_{\pm 1} S_0 + S_0 S_{\pm 1} \right] \,, \notag \\
	& T_{0} = \frac{1}{\sqrt 6} \left[ S_{+1} S_{-1} + S_{-1} S_{+1} + 2 S_0^2 \right] \,.
\end{align} 
Their expectation values $\langle S_M \rangle$ and $\langle T_M \rangle$ are the multipole parameters corresponding to the three polarisation and five alignment components.
The second term in Eq.~(\ref{ec:rhoW}) can be rewritten using the spin operators in the Cartesian basis, and the third one defining the Hermitian operators
\begin{align}
	A_1 = \frac{1}{2} (T_1-T_{-1}) \,, \quad A_2 = {\frac{1}{2 i}} (T_1 + T_{-1}) \,, \notag \\
	B_1 = \frac{1}{2} (T_2+T_{-2}) \,, \quad B_2 = {\frac{1}{2 i}} (T_2 - T_{-2}) \,.
\end{align}
Therefore, the $Z$ boson density matrix elements, parameterised in terms of expectation values of observables, read
\begin{align}
	& \rho_{\pm 1 \pm 1} = \frac{1}{3} \pm \frac{1}{2} \esz + \frac{1}{\sqrt 6} \etz \,,  \notag \\
	& \rho_{\pm 10} = \frac{1}{2\sqrt 2} \left[ \esx \mp i \esy \right] \mp \frac{1}{\sqrt 2} \left[ \eax \mp i \eay \right]  \,, \notag \\
	& \rho_{00} = \frac{1}{3} - \frac{2}{\sqrt 6} \etz \,, \notag \\
	& \rho_{1\, -1} = \ebx - i \eby \,,
	\label{ec:rhoexp}
\end{align}
and $\rho_{m'm}=\rho_{mm'}^*$.
The angular distribution of the $Z$ boson decay products in its rest frame is determinad by $\rho$. Let us restrict ourselves to leptonic decays $Z \to \ell^+ \ell^-$, with  $\ell = e$ or $\ell = \mu$. Using the helicity formalism of Jacob and Wick~\cite{Jacob:1959at}, the amplitude for the decay of a $Z$ boson with third spin component $m$ giving $\ell^-$  with helicity $\lambda_1$ and $\ell^+$ with helicity $\lambda_2$ is written as
\begin{equation}
	\mathcal{M}_{m \lambda_1 \lambda_2} = b_{\lambda_1 \lambda_2} D_{m \lambda}^{1*} (\phi^*,\theta^*,0) \,,
\end{equation}
with $(\theta^*,\phi^*)$ the polar and azimuthal angles of the $\ell^-$ momentum in the $Z$ boson rest frame, $\lambda = \lambda_1 - \lambda_2$ and
\begin{equation}
	D_{m'm}^j (\alpha,\beta,\gamma) =  e^{-i \alpha m'} e^{-i \gamma m} d_{m'm}^j (\beta)
\end{equation}
the so-called Wigner $D$ functions~\cite{wigner};  $b_{\lambda_1 \lambda_2}$ are constants, and all the dependence of the amplitude on the angular variables of the final state products is given by the $D$ functions.
Here, at variance with $W$ boson decays,\footnote{For the $W$ boson the the left-handed interaction fixes $(\lambda_1,\lambda_2) = (\pm 1/2, \mp 1/2)$ for $W^\pm \to \ell^\pm \nu$ decays, that is, there is a single helicity combination in each case.} we have two possible helicity combinations
$(\lambda_1,\lambda_2) = (\pm 1/2, \mp 1/2)$. The differential decay width reads
\begin{eqnarray}
	\frac{d\Gamma}{d\!\cos\theta^* d\phi^*}  & = & C \sum_{m, m' ,\lambda_1 \lambda_2} \rho_{mm'} | b_{\lambda_1 \lambda_2} |^2  e^{i  (m-m')\phi^*}  d_{m\lambda}^{1} (\theta^*) d_{m'\lambda}^{1}(\theta^*) \,,
	\label{ec:dist2d}
\end{eqnarray}
with $\lambda=\lambda_1-\lambda_2 = \pm 1$. The constants $b_{1/2 \, -1/2}$ and $b_{-1/2 \, 1/2}$ are respectively proportional to the the right- and left-handed couplings of the $Z$ boson to the charged leptons $g_R^\ell$, $g_L^\ell$,
\begin{eqnarray}
	b_{1/2 \, -1/2} \;:\;  b_{-1/2 \, 1/2} & = & g_R^\ell \;:\; g_L^\ell \,. 
\end{eqnarray} 
The angular distribution of the $Z$ boson decay products can easily be obtained from the distribution for $W^\pm$ decays~\cite{Aguilar-Saavedra:2015yza} by noting that the only terms that change sign when replacing $\lambda$ by $-\lambda$ are those proportional to $\esi$, $k=1,2,3$. By introducing the polarisation analyser
\begin{equation}
	\eta_\ell = \frac{(g_L^\ell)^2 - (g_R^\ell)^2}{(g_L^\ell)^2 + (g_R^\ell)^2} = \frac{1-4 s_W^2}{1-4 s_W^2 + 8 s_W^4} \,,
	\label{ec:etal}
\end{equation}
with $s_W$ the sine of the weak mixing angle, we get
\begin{align}
	& \frac{1}{\Gamma} \frac{d\Gamma}{d\!\cos\theta^* d\phi^*} = \frac{3}{8\pi} \left\{ 
	\frac{1}{2} (1+\cos^2 \theta^*) -\eta_\ell \esz \cos \theta^*
	+ \left[ \frac{1}{6} - \frac{1}{\sqrt 6} \etz \right] \left( 1-3\cos^2 \theta^* \right)
\right. \notag \\
& ~ -\eta_\ell \esx \cos \phi^* \sin \theta^* -\eta_\ell \esy \sin \phi^* \sin \theta^* 
- \eax \cos \phi^* \sin 2\theta^* - \eay \sin \phi^* \sin 2\theta^* \notag \\
& \left. ~ + \ebx \cos 2 \phi^* \sin^2 \theta^* + \eby \sin 2 \phi^* \sin^2 \theta^* \right\} \,.
\label{ec:distfull}
\end{align}
The angular asymmetries introduced in Ref.~\cite{Aguilar-Saavedra:2015yza} for the measurement of $W$ boson spin observables can be straightforwardly used for the $Z$ boson as well, with the appropriate replacements. We have 
\begin{eqnarray}
	A_\text{FB}^{x'} & = & \frac{1}{\Gamma}  \left[ \Gamma(\cos \phi^* > 0) - \Gamma(\cos \phi^* < 0) \right] = - \frac{3}{4} \eta_\ell \esx \,, \notag \\
	A_\text{FB}^{y'} & = & \frac{1}{\Gamma}  \left[ \Gamma(\sin \phi^* > 0) - \Gamma(\sin \phi^* < 0) \right] =  - \frac{3}{4} \eta_\ell \esy \,, \notag \\
	A_\text{FB}^{z'} & = & \frac{1}{\Gamma} \left[ \Gamma(\cos \theta^* > 0) - \Gamma(\cos \theta^* < 0) \right] 
	= - \frac{3}{4}\eta_\ell  \esz \,, \notag \\
	A_\text{EC}^{z'} & = & \frac{1}{\Gamma} \left[ \Gamma(|\cos \theta^*| > \frac{1}{2} ) - \Gamma(|\cos \theta^*| < \frac{1}{2} ) \right]  = \frac{3}{8}  \sqrt{\frac{3}{2}}  \etz \,, \notag \\
	A_\text{FB}^{x',z'} & = & \frac{1}{\Gamma}  \left[ \Gamma(\cos \phi^* \cos \theta^* > 0) - \Gamma(\cos \phi^* \cos \theta^* < 0) \right] =  - \frac{2}{\pi} \eax \,, \notag \\ 
	A_\text{FB}^{y',z'} & = & \frac{1}{\Gamma}  \left[ \Gamma(\sin \phi^* \cos \theta^* > 0) - \Gamma(\sin \phi^* \cos \theta^* < 0) \right] =  - \frac{2}{\pi} \eay \,, \notag \\
	A_\phi^1 & = & \frac{1}{\Gamma} \left[ \Gamma(\cos 2 \phi^* > 0 ) - \Gamma(\cos 2\phi^* < 0) \right] = \frac{ 2}{\pi} \ebx \,, \notag \\
	A_\phi^2 & = & \frac{1}{\Gamma} \left[ \Gamma(\sin 2 \phi^* > 0) - \Gamma(\sin 2\phi^* < 0) \right] =  \frac{2}{\pi} \eby \,.
	\label{ec:asymlist}
\end{eqnarray}
As seen, these asymmetries separate out the 8 multipole parameters one by one.

\section{Drell-Yan production}
\label{sec:3}

The angular distribution (\ref{ec:distfull}) has been investigated by the CDF~\cite{Aaltonen:2011nr}, CMS~\cite{Khachatryan:2015paa} and ATLAS~\cite{Aad:2016izn} Collaborations in Drell-Yan $Z$ production in hadron collisions, using the Collins-Soper coordinate system~\cite{Collins:1977iv}. The doubly differential angular distribution is parameterised using unknown coefficients labelled as $A_{0-7}$. The density matrix analysis of the $Z$ boson decay provides an interpretation of the measured coefficients  in terms of $Z$ boson spin observables,
\begin{align}
	& A_0 = \frac{2}{3} - 2 \sqrt{\frac{2}{3}} \etz \,, && A_1 = -2 \eax \,, \notag \\[1mm]
	& A_2 = 4 \ebx \,, && A_3 = -2 \eta_\ell \esx \,, \notag \\[1mm]
	& A_4 = -2 \eta_\ell \esz \,, && A_5 = 2 \eby \,, \notag \\[1mm]
	& A_6 = -2 \eay \,, && A_7 = -2 \eta_\ell \esy \,.
\end{align}
Experiments have measured these coefficients differentially, as a function of the transverse momentum and rapidity of the $Z$ boson. An interpretation in terms of $Z$ spin observables, apart from providing more insight into the nature of the physical observables measured, provides a rationale for the smallness of $A_3$, $A_4$ and $A_7$, since they are proportional to the small polarisation analyser $\eta_\ell \simeq 0.14$. The most recent measurement by the ATLAS Collaboration~\cite{Aad:2016izn} exhibits a noticeable deviation in $A_2$ with respect to next-to-leading order~\cite{Alioli:2008gx} and next-to-next-to-leading order~\cite{Catani:2009sm} SM predictions. Nevertheless, the CMS Collaboration finds agreement with the multi-leg SM prediction from {\scshape MadGraph5\_aMC@NLO}~\cite{Alwall:2014hca}. This coefficient corresponds to the rank-two alignment $\ebx$.

We also point out that --- besides the method commonly used to extract the angular coefficients $A_{0-7}$ based on integration with suitable weight functions~\cite{Mirkes:1994eb} --- the asymmetries in (\ref{ec:asymlist}) provide an alternative way for their determination. Whether this simpler method also gives more precise results depends on the systematic uncertainties in each case, and a detailed analysis is compulsory to draw any conclusion.

\section{$Z$ boson plus $\met$ production}
\label{sec:4}

The production at the LHC of final states containing a same-flavour opposite-sign lepton (electron or muon) pair with invariant mass around the $Z$ boson mass, possibly jets, and large $\met$ is very relevant in the search by ATLAS~\cite{ATLASnew} and CMS~\cite{CMSnew} of Supersymmetry (SUSY) signals and collider production of Dark Matter~\cite{Bell:2012rg}.
Besides the simple event counting, the use of spin observables in these $Z+\met$ searches
provides an additional handle to test the SM predictions and uncover possible effects of new physics.

Similarly to the previous section, the full angular distribution (\ref{ec:distfull}) can be measured differentially as a function of the $\met$ in these final states. The leading SM processes yielding $Z$ plus missing energy are (i) $ZZ$ production, with $ZZ \to \ell^+ \ell^- \nu \bar \nu$; (ii) $WZ$ production, with $Z \to \ell^+ \ell^-$, $W \to \ell' \nu$, and the additional charged lepton $\ell'$ undetected, because of having a small transverse momentum or large rapidity. We have used {\scshape MadGraph5\_aMC@NLO} to simulate these processes at the tree level, followed by hadronisation by {\scshape Pythia}~\cite{Sjostrand:2007gs}, in order to estimate the SM prediction for the $Z+\met$ final state in $pp$ collisions at a CM energy of 13 TeV. We restrict our analysis to events happening at the $Z$ peak, with the two charged leptons in an invariant mass window of $88-94$ GeV.

\begin{figure}[tb]
	\begin{center}
		\includegraphics[height=6cm,clip=]{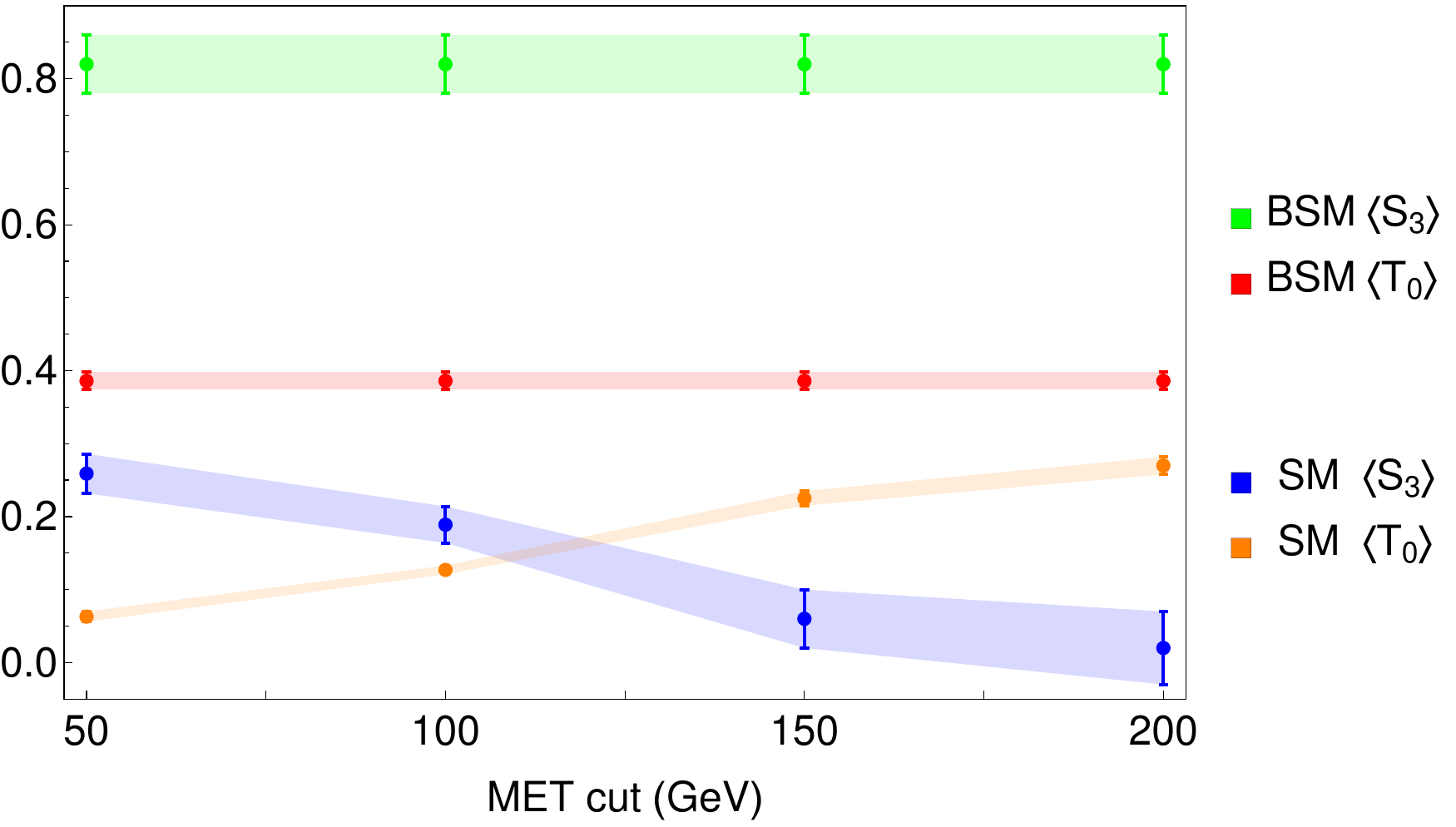}
		\caption{SM predictions for $\esz$ and $\etz$ for $Z+\met$ final states, as a function of the lower cut on $\met$.
		Above them, the same result from the BSM model described in the text.}
		\label{fig:met}
	\end{center}
\end{figure}

We set our reference system in the $Z$ boson rest frame with the $\hat z^\prime$ axis in its momentum
direction, while the other two axes are left unspecified --- so the  non-diagonal elements of the density
matrix vanish, leaving only $\esz$ and $\etz$ as observables. The calculated values of
$\esz$ and $\etz$, as a function of the lower cut on $\met$, are presented in Fig.~\ref{fig:met}. The bands represent the Monte Carlo statistical uncertainty of our results. Notice that our simulation only includes signal events, so the meaning of these MET cuts is dynamical, i.e. the inclusion of a MET threshold leads to a different production mechanism of the $Z$ boson. For the SM predictions (blue and orange bands), the large dependence of these two observables on the $\met$ cut makes their measurement very interesting to test the SM, as well as providing a reference for beyond the Standard Model (BSM) searches. 

\begin{figure}[htb]
\begin{center}
\begin{tabular}{ccc}
\includegraphics[height=2.8cm,clip=]{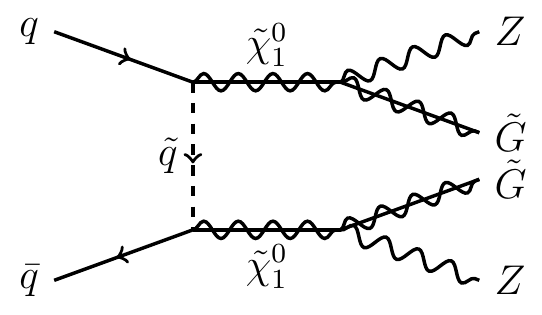} & \quad \quad &
\includegraphics[height=3.2cm,clip=]{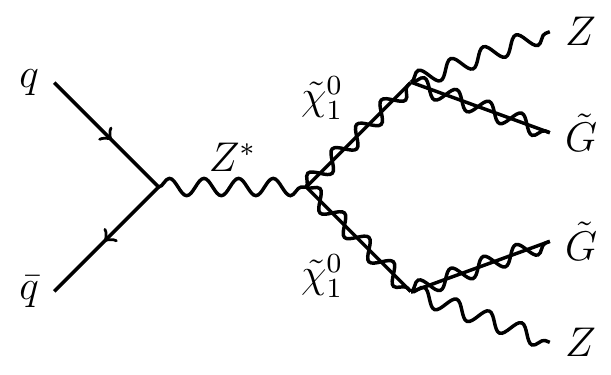}
\end{tabular}
\end{center}
	\caption{Neutralino (decaying to $Z$ plus gravitino) production mechanisms considered in the SUSY dark matter model.}
	\label{fig:BSM}
\end{figure}

To illustrate the power of the declared strategy, we compare with the expected values of these observables
in a SUSY dark matter model with the gravitino $\tilde G$ as lightest supersymmetric particle (LSP) and the lightest neutralino $\tilde \chi^0_1$ as next-to-LSP. 
We consider a massless gravitino and a lightest neutralino $\chi_1^0$ whose mass is around 100 GeV.
For simplicity, we assume the direct electroweak production of a pair $\tilde \chi^0_1 \tilde \chi^0_1$ from quark-antiquark, as shown in Fig.~\ref{fig:BSM}.
The simulation is performed within the {\scshape MadGraph5\_aMC@NLO} framework utilising the gravitino implemented in \textsc{FeynRules} output in the Universal FeynRules Output (UFO)~\cite{gravitino-ufo}. The BSM values of $\etz$ and $\esz$ do not depend on the MET threshold. The former is $\etz = 1/\sqrt{6}$ due to angular momentum conservation, as discussed in the next section (with $j=1/2$, $j^\prime = 3/2$).  The latter is fixed for a particular process, but may change in the presence of an additional production channel. Since in both diagrams in Fig.~\ref{fig:BSM} we have the same incoming and outgoing particles, $q \bar q \to \chi_1^0 \chi_1^0 \to Z \tilde G Z \tilde G$, the kinematical distributions, including the missing transverse momentum, should be the same. Therefore, the change in the MET threshold should not affect the relative contribution of each diagram, and $\esz$ should be independent of the MET cut.

The consideration of Z boson plus MET production in SUSY models with gravitinos is
motivated by previous work~\cite{zmet-excess}, where it was shown that Gauge Mediation models could lead to a privileged new production mechanism of $Z$ bosons. In this model, the production mechanism of $Z$ bosons plus $\met$ is $\tilde \chi^0 \to Z\; \tilde G$,
which produces $Z$ bosons with the diagonal spin parameters shown in Fig.~\ref{fig:met} (green and red bands).

\section{$Z$ bosons from heavy particle decays}
\label{sec:5}

Let us consider that a $Z$ boson is produced in the two-body decay of some spin-$j$ particle $A$, yielding also a spin-$j'$ particle $B$ as decay product,
\begin{equation}
	A(j,m) \to Z(1,\lambda_1) B(j',\lambda_2) \,,
\end{equation}
where $\lambda_1$, $\lambda_2$ are the helicities of $Z$ and $B$ in the rest frame of the decaying particle $A$, and $m$ its third spin component. Let us fix a $(\hat x,\hat y,\hat z)$ coordinate system in the rest frame of $A$. The amplitude for the decay can be written as
\begin{equation}
	\mathcal{M}_{m \lambda_1 \lambda_2} = a_{\lambda_1 \lambda_2} D_{m \lambda}^{j*} (\phi,\theta,0) \,,
\end{equation}
with $\theta$ and $\phi$ the polar and azimuthal angle of the $Z$ boson momentum, in full analogy to the $Z$ boson decay discussed in Section~\ref{sec:2}. Restricting ourselves to particles up to spin $3/2$, we have six combinations for $j$ and $j'$, collected in Table~\ref{tab:jjprime}. We assume that spin $3/2$ particles are massless, so their only possible helicities are $\pm 3/2$; furthermore, the decaying particle $A$ cannot have spin $3/2$ in this case. For each pair $j,j'$, angular momentum conservation implies that only a subset of $a_{\lambda_1 \lambda_2}$ are non-zero; these combinations are also given in Table~\ref{tab:jjprime}.

\begin{table}[t]
	\caption{Non-zero amplitudes for $A(j,m) \to Z(1,\lambda_1) B(j',\lambda_2)$ with arbitrary spins.}
	\label{tab:jjprime}
	\begin{center}
		\begin{tabular}{ccccl}
			$j$ &~& $j'$ &~& Non-zero amplitudes\\
			\hline
			$1/2$ & & $3/2$ & & $a_{1\,3/2}$, $a_{-1\,-3/2}$ \\
			$1/2$ &  &$1/2$ & & $a_{1\,1/2}$, $a_{0\,1/2}$, $a_{0\,-1/2}$, $a_{-1\,-1/2}$ \\
			$1$    &  & $1$   & & $a_{1 \, 1}$, $a_{1 \, 0}$, $a_{0 \, 1}$, $a_{0 \, 0}$, $a_{0 \, -1}$, $a_{-1\, 0}$, $a_{-1 \, -1}$ \\
			$1$    &  & $0$   & & $a_{1 \, 0}$, $a_{0 \, 0}$, $a_{-1\, 0}$ \\
			$0$    &  & $1$   & & $a_{1 \, 1}$, $a_{0 \, 0}$, $a_{-1\, -1}$ \\
			$0$    & & $0$    & & $a_{0 \, 0}$
		\end{tabular}
	\end{center}
\end{table}

Angular momentum conservation restricts the form of the $Z$ boson spin density matrix, whose elements also depend on the angles $\theta$ and $\phi$. Differential measurements of the $Z$ spin observables as functions of $\theta$ and $\phi$ are possible with sufficient statistics, but for simplicity we consider here the integrated density matrix, using for the $Z$ boson rest frame the $(\hat x',\hat y',\hat z')$ coordinate system implied by the standard boost from the rest frame of $A$, with the $\hat z'$ axis in the $Z$ helicity direction.\footnote{This boost is given by a rotation $R(\phi,\theta,0)$ in the Euler parameterisation, followed by a pure boost in the $Z$ momentum direction to set it at rest.} Experimentally, this would require the measurement of the momentum of the $B$ particle too, in order to reconstruct the momentum of $A$ and choose a coordinate system in its rest frame. (In the previous section we have not made such assumption.) With this setup, integrating over $\theta$ and $\phi$ does not generally give a diagonal $\rho$, and a dependence on the values of $\esz$ and $\etz$ for the particle $A$, denoted here as $\eszA$ and $\etzA$, respectively,  is retained. This subtle effect is due to the fact that, when performing the above specified standard boost from the rest frame of $A$, the $\hat y'$ axis is always in the $xy$ plane (see for example Ref.~\cite{libro} for a detailed discussion) irrespectively of the values of $\theta$ and $\phi$, and the integration over these two angles is not equivalent to considering an isotropic distribution of the $\hat x'$ and $\hat y'$ axes.

The predictions for the different combinations of $j,j'$ are as follows.
\begin{itemize}
	\item $j=1/2$, $j' = 3/2$. The density matrix has all entries vanishing  except $\rho_{11}$ and $\rho_{-1 -1}$, implying $\esx = \esy = 0$, $\eax = \eay = 0$, $\ebx = \eby = 0$. The only non-trivial observables are
		\begin{eqnarray}
			\esz  =  \left[  |a_{1 \, 3/2}|^2 - |a_{-1 \, -3/2}|^2 \right] / \mathcal{N} \,,\hspace{2cm}
			\etz  =  1/\sqrt 6\,.
		\end{eqnarray}
		Here and below, $\mathcal{N}$ is the sum of the moduli squared of all the non-zero amplitudes in Table~\ref{tab:jjprime} for the case under consideration.
		These results were referred to in Section 4 of this paper.
	\item $j=1/2$, $j' = 1/2$. Here $\rho_{1 -1} = 0$, therefore $\ebx = \eby = 0$. The remaining observables can be non-zero,
		\begin{eqnarray}
			\esz & = & \left[  |a_{1 \, 1/2}|^2 - |a_{-1 \, -1/2}|^2 \right] / \mathcal{N} \,, \notag \\
			\etz & = & \frac{1}{\sqrt 6} \left\{ 1 - 3 \left[ |a_{0 \, 1/2}|^2 + |a_{0 \, -1/2}|^2 \right] /\mathcal{N} \right\} \,, \notag \\
			\esx & = & -\frac{\pi}{\sqrt 2} \eszA \Re [ a_{-1 \,-1/2} \, a_{0 \,-1/2}^*  + a_{1 \,1/2} \, a_{0 \,1/2}^* ] / \mathcal{N}  \,, \notag \\
			\esy & = & -\frac{\pi}{\sqrt 2} \eszA \Im [ a_{-1 \,-1/2} \, a_{0 \,-1/2}^*  - a_{1 \,1/2} \, a_{0 \,1/2}^* ] / \mathcal{N}  \,, \notag \\
			\eax & = & -\frac{\pi}{2\sqrt 2} \eszA \Re [ a_{-1 \,-1/2} \, a_{0 \,-1/2}^*  - a_{1 \,1/2} \, a_{0 \,1/2}^* ] / \mathcal{N}  \,, \notag \\
			\eay & = & -\frac{\pi}{2\sqrt 2} \eszA \Im [ a_{-1 \,-1/2} \, a_{0 \,-1/2}^* + a_{1\, 1/2} \, a_{0 \, 1/2}^*  ] / \mathcal{N}  \,.
		\end{eqnarray}
	\item $j=1$, $j'=0,1$. In these two cases all the density matrix elements and $Z$ boson spin observables are generally different from zero. We can write them as
		\begin{eqnarray}
			\esz & = & \left[ C_{1} - C_{-1} \right] / \mathcal{N} \,, \notag \\
			\etz & = & \frac{1}{\sqrt 6} \left[ 1 - 3 C_0 /\mathcal{N} \right] \,, \notag \\
			\esx & = & -\frac{3\pi}{8} \eszA \Re [ C_{-10} + C_{10} ] / \mathcal{N} \,, \notag \\ 
			\esy & = & -\frac{3\pi}{8} \eszA \Im [ C_{-10} - C_{10} ] / \mathcal{N} \,, \notag \\ 
			\eax & = & -\frac{3\pi}{16} \eszA \Re [ C_{-10} - C_{10} ] / \mathcal{N} \,, \notag \\
			\eay & = & -\frac{3\pi}{16} \eszA \Im [ C_{-10} + C_{10} ] / \mathcal{N} \,, \notag \\
			\ebx & = & \sqrt{\frac{3}{2}} \etzA \Re C_2 \,, \notag \\
			\eby & = & \sqrt{\frac{3}{2}} \etzA \Im C_2 \,,
		\end{eqnarray}
		where we have abbreviated products of amplitudes $a_{\lambda_1 \lambda_2}$ for $j'= 0$ ($j' = 1$) as
		\begin{eqnarray}
			C_{1}  & = & |a_{1\,0}|^2 \;( \;+  \;|a_{1\,1}|^2 \;) \,, \notag \\
			C_{-1} & = & |a_{-1\,0}|^2 \;( \;+  \;|a_{-1\,-1}|^2 \;) \,, \notag \\
			C_{0} & = & |a_{0\,0}|^2 \;( \;+  \;|a_{0 \,1}|^2 + |a_{0 \,-1}|^2 \;) \,, \notag \\
			C_{10} & = & a_{1\,0} a_{0\,0}^* \;( \;+ a_{1\,1} \, a_{0\,1}^* \;) \,, \notag \\
			C_{-10} & = & a_{-1\,0} \, a_{0\,0}^* \;( \;+ a_{-1\,-1} \, a_{0\,-1}^* \;) \,, \notag \\
			C_2 & = & a_{-1 \, 0} \, a_{1 \, 0}^*  \,.
		\end{eqnarray}
	\item $j=0$, $j' = 1$. The $Z$ spin density matrix is diagonal, implying $\esx = \esy = 0$, $\eax = \eay = 0$, $\ebx = \eby = 0$. The diagonal spin observables are
		\begin{eqnarray}
			\esz & = & \left[  |a_{1 \, 1}|^2 - |a_{-1 \, -1}|^2 \right] / \mathcal{N} \,,  \notag \\
			\etz & = & \frac{1}{\sqrt 6} \left[ 1 - 3 |a_{0 \, 0}|^2 /\mathcal{N} \right] \,.
		\end{eqnarray}
	\item $j=0$, $j' = 0$. This case is particularly interesting, implying a full longitudinal $Z$ (helicity $\lambda=0$) or,
		equivalently, a $p$-wave (orbital angular momentum $l=1$) state. The only non-zero density matrix element is $\rho_{00} = 1$, therefore
		 $\esx = \esy = \esz = 0$, $\eax = \eay = 0$, $\ebx = \eby = 0$ and
		 \begin{eqnarray}
			\etz & = & -\frac{2}{\sqrt{6}}\,.	
		 \end{eqnarray}
		 It would be the situation in the decay $A(0^-)\to Z + h(0^+)$ in the Higgs sector of SUSY and left-right models.

\end{itemize}
This analysis can be applied to current and future heavy resonance searches. If a new resonance is discovered, the use of spin observables in the $Z$ leptonic decay products might shed light into its nature. If any spin observable is measured to have a non-trivial value, or $\etz \neq -2/\sqrt{6}$, the $j=0$, $j' = 0$ hypothesis is discarded. If, additionally, a non-zero value is found for any off-diagonal spin observable, the $j=0$, $j'=1$ hypothesis can be discarded too. Furthermore, a non-trivial value for $\left< B_{1,2} \right>$ can only be explained by a $j=1$ parent.
However, discriminating the two $j=1$ possibilities requires an analysis of the decay of the extra particle $B$.

To conclude this section we remark that, as anticipated, the classification and relation of spin observables with decay amplitudes obtained also apply for a $W$ boson, namely to the decay
\begin{equation}
	A (j,m) \to W(1,\lambda_1) B(j',\lambda_2)
\end{equation}
because we have not used any other property than the spin. In particular, one example of $j=1/2$, $j'=1/2$ decays is given by $t \to W b$, studied in Ref.~\cite{Aguilar-Saavedra:2015yza}.

\section{Conclusions} 
\label{sec:6}

With the wealth of collision data being accumulated by LHC experiments, the measurement of the eight $Z$ boson spin observables becomes feasible from the angular distribution of its leptonic decay.
We have proved in this work the discriminating power of these polarisation and alignment observables
for identifying the production mechanism of the $Z$ boson, with apparent different values for known processes in the SM and for extended models. These observable quantities thus play the role of messengers from the hidden physics leading to the $Z$ boson production.

Interesting physical processes include the Drell-Yan $Z$ boson production, for which we have given the physical interpretation of the parameters of the lepton angular distribution measured by ATLAS and CMS,
pointing out the alternative method of extracting the $Z$ boson spin observables by means of selected asymmetries. When we move to processes able to generate large missing transverse energy, the SM reference values for the $Z$ boson longitudinal polarisation $\esz$ and alignment $\etz$ present a characteristic rapid variation above 100 GeV of MET, contrary to the values and behaviour obtained from SUSY models with new sources of $Z$ boson production like the decay of neutralinos to gravitinos. For two-body decays of a heavy particle involving a $Z$ boson (or $W$ boson) in its decay products, we have demonstrated that different spin assignments of the parent and daughter particles lead to specific zeros and values of the $Z$ boson polarisations and alignments. The use of these observables will increasingly become an invaluable interesting handle to test the SM predictions and look for new physics.

\section*{Acknowledgements}
This research has been supported by MINECO Projects  FPA 2016-78220-C3-1-P, FPA 2015-65652-C4-1-R, 
FPA 2014-54459-P, FPA 2013-47836-C3-2-P (including ERDF),
Junta de Andaluc\'{\i}a Project FQM-101,
Generalitat Valenciana Project GV PROMETEO II 2013-017,
Severo Ochoa Excellence Centre Project SEV 2014-0398, and European Commission through the contract PITN-GA-2012-316704 (HIGGSTOOLS).
VAM acknowledges the support by Spanish National Research Council (CSIC) under the CT Incorporation Program 201650I002, and
AS acknowledges the MECD support through the FPU14/04678 grant.


\begin{thebibliography}{99}

	\bibitem{ref:LEP}
	S.~Schael {\it et al.} [ALEPH, DELPHI, L3, OPAL, SLD, LEP Electroweak Working Group, SLD Electroweak and Heavy Flavour Groups],
	Phys. Reports 427 (2006) 257;
	for updates see http://lepewwg.web.cern.ch/LEPEWWG/

\bibitem{Groote:1995ky}
  S.~Groote and J.~G.~Korner,
  Z.\ Phys.\ C {\bf 72} (1996) 255
   Erratum: [Z.\ Phys.\ C {\bf 70} (2010) no.1-2,  531]
  [hep-ph/9508399].

\bibitem{Groote:1996nc}
  S.~Groote, J.~G.~Korner and M.~M.~Tung,
  Z.\ Phys.\ C {\bf 74} (1997) 615
  [hep-ph/9601313].

\bibitem{Tung:1996dq}
  M.~M.~Tung, J.~Bernab\'eu and J.~Pe\~narrocha,
  Nucl.\ Phys.\ B {\bf 470} (1996) 41
  [hep-ph/9601277].

\bibitem{Fischer:2001gp}
  M.~Fischer, S.~Groote, J.~G.~Korner and M.~C.~Mauser,
  Phys.\ Rev.\ D {\bf 65} (2002) 054036
  [hep-ph/0101322].

\bibitem{Gounaris:1992kp} 
G.~Gounaris, J.~Layssac, G.~Moultaka and F.~M.~Renard,
Int.\ J.\ Mod.\ Phys.\ A {\bf 8}, 3285 (1993).



	\bibitem{Mirkes:1994eb} 
	E.~Mirkes and J.~Ohnemus,
	Phys.\ Rev.\ D {\bf 50}, 5692 (1994)
	[hep-ph/9406381].

	\bibitem{Aguilar-Saavedra:2015yza} 
	J.~A.~Aguilar-Saavedra and J.~Bernab\'eu,
	Phys.\ Rev.\ D {\bf 93}, 011301 (2016)
	[arXiv:1508.04592 [hep-ph]].

	\bibitem{Rahaman:2016pqj}
	R.~Rahaman and R.~K.~Singh,
	Eur.\ Phys.\ J.\ C {\bf 76} (2016) no.10,  539
	[arXiv:1604.06677 [hep-ph]].

	\bibitem{Jacob:1959at} 
	M.~Jacob and G.~C.~Wick,
	Annals Phys.\  {\bf 7}, 404 (1959)
	[Annals Phys.\  {\bf 281}, 774 (2000)].

	\bibitem{wigner}
	E. P. Wigner, {\em Group Theory and Its Application to the Quantum Mechanics of Atomic Spectra}, edited by J. J. Griffin (Academic Press, New York, 1959).

	\bibitem{Aaltonen:2011nr} 
	T.~Aaltonen {\it et al.} [CDF Collaboration],
	Phys.\ Rev.\ Lett.\  {\bf 106}, 241801 (2011)
	[arXiv:1103.5699 [hep-ex]].

	\bibitem{Khachatryan:2015paa} 
	V.~Khachatryan {\it et al.} [CMS Collaboration],
	Phys.\ Lett.\ B {\bf 750}, 154 (2015)
	[arXiv:1504.03512 [hep-ex]].

	\bibitem{Aad:2016izn} 
	G.~Aad {\it et al.} [ATLAS Collaboration],
	JHEP {\bf 1608} (2016) 159
	[arXiv:1606.00689 [hep-ex]].

	\bibitem{Collins:1977iv}
	J.~C.~Collins and D.~E.~Soper,
	Phys.\ Rev.\ D {\bf 16} (1977) 2219.

	\bibitem{Alioli:2008gx} 
	S.~Alioli, P.~Nason, C.~Oleari and E.~Re,
	JHEP {\bf 0807}, 060 (2008)
	[arXiv:0805.4802 [hep-ph]].

	\bibitem{Catani:2009sm} 
	S.~Catani, L.~Cieri, G.~Ferrera, D.~de Florian and M.~Grazzini,
	Phys.\ Rev.\ Lett.\  {\bf 103}, 082001 (2009)
	[arXiv:0903.2120 [hep-ph]].

	\bibitem{Alwall:2014hca}
	J.~Alwall {\it et al.},
	JHEP {\bf 1407} (2014) 079
	[arXiv:1405.0301 [hep-ph]].

	\bibitem{ATLASnew}
	M.~Aaboud {\it et al.} [ATLAS Collaboration],
	CERN-PH-2016-260,
	[arXiv:1611.05791 [hep-ex]].

	\bibitem{CMSnew}
	V.~Khachatryam {\it et al.} [CMS Collaboration],
	JHEP 1612 (2016) 013,
	[arXiv:1607.00915 [hep-ex]].

	\bibitem{Bell:2012rg}
	N.~F.~Bell, J.~B.~Dent, A.~J.~Galea, T.~D.~Jacques, L.~M.~Krauss and T.~J.~Weiler,
	Phys.\ Rev.\ D {\bf 86} (2012) 096011
	[arXiv:1209.0231 [hep-ph]].

	\bibitem{Sjostrand:2007gs}
	T.~Sjostrand, S.~Mrenna and P.~Z.~Skands,
	Comput.\ Phys.\ Commun.\  {\bf 178} (2008) 852
	[arXiv:0710.3820 [hep-ph]].

	\bibitem{gravitino-ufo}
	N.~D.~Christensen {\it et al.},
	Eur.\ Phys.\ J.\ C {\bf 73} (2013) no.10,  2580
	[arXiv:1308.1668 [hep-ph]].

	\bibitem{zmet-excess}
	G.~Barenboim, J.~Bernabeu, V.~A.~Mitsou, E.~Romero and O.~Vives,
	Eur.\ Phys.\ J.\ C {\bf 76} (2016) 57
	[arXiv:1503.04184 [hep-ph]].


	\bibitem{libro}
	J. A. Aguilar-Saavedra, {\it Helicity formalism and applications} (Godel, Granada, 2015).







\end{thebibliography}
\end{document}